\documentstyle[aps,epsfig]{revtex}

\tighten
\begin{document}

\draft

\twocolumn[\columnwidth\textwidth\csname@twocolumnfalse\endcsname
\title{\bf $E1-E2$ interference in the Coulomb dissociation of $^8$B}

\author{P. Banerjee and R. Shyam}
\address{Theory Group, Saha Institute of Nuclear Physics,
\\1/AF Bidhan Nagar, Calcutta - 700 064, INDIA}
\date{\today}

\maketitle

\begin{abstract}
We investigate the effects arising out of the $E1 - E2$ interference in
the Coulomb dissociation of $^8$B at beam energies below and around
50 MeV/nucleon. The theory has been formulated within a first order 
semiclassical scheme of Coulomb excitation, in which both the ground state 
and the continuum state wave functions of $^8$B enter as inputs. We find that
the magnitude of the interference could be large in some cases. However,
there are some specific observables which are free from the effects of 
the $E1 - E2$ interference, which is independent of the models used to describe
the structure of $^8$B. This will be useful for the analysis of the breakup 
data in relation to the extraction of the astrophysical factor $S_{17}(0)$.
\end{abstract}
\addvspace{5mm}]
\vfill
\section{\bf INTRODUCTION}
The Coulomb dissociation (CD) method provides an alternate indirect way to 
determine the cross sections for the radiative capture reaction $^7$Be (p,
$\gamma$) $^8$B at low relative energies \cite{bau}, which is a key reaction 
of the $p-p$ chain through which energy is generated in the sun.
The rate of this reaction is the most uncertain nuclear input to
the standard solar model calculations \cite{bah}, which affects the high energy
solar neutrino flux and bears significantly on the solar neutrino problem. The 
CD method reverses the radiative capture by the dissociation of a projectile 
(the fused system) in the Coulomb field of 
a target, by making the assumption that nuclei do not interact strongly and the 
electromagnetic excitation process is dominated by a single multipolarity
\cite{bau}.
Therefore, one has to pin down a kinematical domain where breakup cross sections
due to different multipolarities ($E1$ and $E2$) can be clearly separated from 
each other. At the same time, it is also necessary to investigate the 
importance of the $E1 - E2$ interference effects which has not been considered 
so far in most of the analyses of the Coulomb dissociation data of $^8$B.
 
The magnitudes of the $E1 - E2$ interference cross sections could be 
appreciable for certain observables in the Coulomb dissociation of $^8$B. It 
was shown in Refs.~\cite{esb,bert} that the dissociation of $^8$B 
around 45 MeV/nucleon beam energy is dominated by $E1$ transitions but the 
interference with $E2$ amplitudes produces large asymmetries in the angular 
and momentum distributions of the breakup fragments. This fact has been used 
in a recent measurement of the parallel momentum distribution of the $^7$Be 
fragment resulting from the breakup of $^8$B on a Pb target at 44 MeV/nucleon 
beam energy to put constraint on the contributions of the $E2$ component in 
the breakup of $^8$B \cite{dav}. 

In this paper, our aim is to investigate the role of the $E1 - E2$ interference
effects in the analysis of the Coulomb dissociation data 
at beam energies 25.8 MeV at Notre Dame \cite{von} and around 50 MeV/nucleon
measured at RIKEN \cite{mot,kik}. We would like to examine 
the importance of the interference effects in different cases and try to see 
whether 
they could be really absent in the above data. To this end, we have developed 
a first order semiclassical theory for the electromagnetic excitation of a 
composite nucleus leading to continuum final states which allows to 
estimate the $E1$ and $E2$ breakup contributions separately as 
well as their interference. Our formulation is different from that of 
Refs.~\cite{esb,bert} in the sense that we do not assume straight line 
trajectories for the motion of the projectile in the Coulomb field of the
target nucleus. Our theory is closer to that
of Baur and Weber \cite{baw} where triple differential cross sections with
respect to the relative and centre of mass (c.m.) angles and energies of the
fragments are determined. However, we have used a more general coupling scheme 
of angular momenta and spins than those in Ref.~\cite{baw}.

We organize the paper as follows. The theoretical formalism has been described 
in section 2. We describe the structure model adopted for $^8$B in section 3.
The results and discussions on them are presented in section 4. Finally, the
summary and conclusions are given in section 5.

\section{\bf FORMALISM}

We consider the following reaction
\begin{eqnarray}
a + A \rightarrow c + v + A,
\end{eqnarray}
in which the composite projectile $a (= c + v)$ breaks up into a core $c$ and 
a valence
particle $v$ in the Coulomb field of the target nucleus $A$ which remains in
the ground state (elastic breakup). We consider
electric transitions of dipole and quadrupole types only. It is to be noted 
that the $M1$ transition contribution is small for the reactions considered 
here at beam energies $<$52 MeV/nucleon \cite{rs}.

In the initial channel, for given spins $s_{c_i}$ and $s_{v_i}$ of the core 
and valence particles, the projectile wave function is
\begin{eqnarray}
\psi _{I_iM_i}({\vec r})&=&\sum _{l_iL_i}\left[[Y_{l_i}(\hat{\vec r})\otimes 
\chi_{s_{v_i}}
]_{L_i}\otimes \chi_{s_{c_i}}\right]_{I_iM_i}\nonumber \\
&\times&{f_{(L_is_{c_i})I_i}(r)\over r}.
\end{eqnarray}
In the above, $l_i$ is the orbital angular momentum of the valence particle
relative to the core and $I_i$ is the total spin of the projectile in its 
ground state with projection $M_i$. $\chi$'s are the spin wave functions of the
core and the valence particle. 

More explicitly
\begin{eqnarray}
\psi _{I_iM_i}({\vec r})&=&[\sum _{{l_im_{l_i}\atop L_iM_{L_i}}\atop m_{v_i}m_{c_i}}
(-1)^{l_i + s_{v_i} - M_{L_i}} \sqrt{2L_i + 1}\nonumber \\&\times&
\left(\begin{array}{ccc}l_i&s_{v_i}&L_i\\m_{l_i}&m_{v_i}&-M_{L_i}\end{array}
\right)Y_{l_im_{l_i}}(\hat {\vec r})\chi _{s_{v_i}m_{v_i}}]\nonumber 
\\&\times& (-1)^{L_i + s_{c_i} - M_i} 
\sqrt{2I_i + 1}
\left(\begin{array}{ccc}L_i&s_{c_i}&I_i\\M_{L_i}&m_{c_i}&-M_i\end{array}\right)
\nonumber \\&\times& 
\chi _{s_{c_i}m_{c_i}}
{1\over r}f_{(L_is_{c_i})I_i}(r).
\end{eqnarray}
In the above, $m_{l_i}$ is the projection of $l_i$.

For the final channel, the wave function is
\begin{eqnarray}
&&\psi _{\vec{k}s_{v_f}m_{v_f}s_{c_f}m_{c_f}}(\vec{r})=\sum _{{I_fM_fm_{v_f}'
m_{c_f}'\atop l_fm_{l_f}m_{l_f}'}\atop L_fM_{L_f}M_{L_f}'}[[
(-1)^{l_f + s_{v_f} - M_{L_f}'}
\nonumber \\ &\times&
\sqrt{2L_f + 1}
\left(\begin{array}{ccc}l_f&s_{v_f}&L_f\\m_{l_f}'&m_{v_f}'&-M_{L_f}'\end{array}\right)
Y_{l_fm_{l_f}'}(\hat {\vec r})
\chi _{s_{v_f}m_{v_f}'}]
\nonumber \\&\times&
(-1)^{L_f + s_{c_f} - M_f}\sqrt{2I_f + 1}
\left(\begin{array}{ccc}L_f&s_{c_f}&I_f\\M_{L_f}'&m_{c_f}'&-M_f\end{array}
\right)
\nonumber \\&\times&
\chi _{s_{c_f}m_{c_f}'}]
(-1)^{l_f + s_{v_f} - M_{L_f}}\sqrt{2L_f + 1}\nonumber \\&\times&
\left(\begin{array}{ccc}l_f&s_{v_f}&L_f\\m_{l_f}&m_{v_f}&-M_{L_f}\end{array}\right)
(-1)^{L_f + s_{c_f} - M_f}\sqrt{2I_f + 1}
\nonumber \\&\times&
\left(\begin{array}{ccc}L_f&s_{c_f}&I_f\\M_{L_f}&m_{c_f}&-M_f\end{array}\right)
Y^{\star}_{l_fm_{l_f}}(\hat {\vec k})
{1\over r}g_{(L_fs_{c_f})I_f}(k,r),
\end{eqnarray}
where $\vec k$ is the wave vector associated with the relative motion between
the two fragments in the continuum. $l_f$ is the orbital angular momentum for
the core-valence relative motion and $I_f$ is the total spin of the $c + v$ 
system in the final channel.

We assume that the electromagnetic multipole operators act only
on the relative motion variables of the system. This leads to the following
conditions: $s_{c_i}=s_{c_f}=s_c$, $s_{v_i}=s_{v_f}=s_v$ and $m_{c_i}=m_{c_f}'$,
$m_{v_i}=m_{v_f}'$.

Now we define a reduced matrix element of the multipole operator $M(El,m)
(=[Z_ce({A_v\over A_a})^l + (-1)^lZ_ve({A_c\over A_a})^l]r^l Y_{lm}(\hat 
{\vec r})$ with multipolarity $l$ having projection $m$) between the final
and the initial state:
\begin{eqnarray}
&&\langle(L_is_c)I_i\mid\mid M(El)\mid\mid (L_fs_c)I_f\rangle =
(-1)^{l_i} 
\nonumber \\&\times&
\sqrt{(2l_i + 1)(2l_f + 1)(2l + 1)\over 4\pi}
\left(\begin{array}{ccc}l_i&l&l_f\\0&0&0\end{array}\right)
\nonumber \\&\times&
[Z_ce({A_v\over A_a})^l + (-1)^lZ_ve({A_c\over A_a})^l]
\nonumber \\&\times&
\int ^{\infty}_0dr f^{\star}_{(L_is_c)I_i}(r).r^l.g_{(L_fs_c)I_f}(k,r).
\end{eqnarray}
In Eq.(5), $Z_i (i = c, v)$ is the charge of the $i$th fragment and $A_j
(j = c, v, A)$ is the mass number of the $j$th partcle. 

With this definition, the matrix element of the multipole operator is
given by
\begin{eqnarray}
&&\langle\psi _{I_iM_i}\mid M(El, m)\mid \psi _{\vec{k}s_{v_f}m_{v_f}s_{c_f}
m_{c_f}}\rangle\nonumber \\
&=& \sum_{{l_im_{l_i}\atop L_im_{L_i}}\atop m_{v_i}m_{c_i}}
\sum_{{I_fM_f\atop l_fm_{l_f}m_{l_f}'}\atop L_fM_{L_f}M_{L_f}'}
(-1)^{-m_{v_i} + L_i + 2L_f + 3s_v + 3s_c - M_i - M_{L_f}} 
\nonumber \\ &\times& 
(-1)^{- M_{L_f}' - 2M_f}\sqrt{(2L_i + 1)(2I_i + 1)}(2L_f + 1)
\nonumber \\ &\times&
(2I_f + 1)
\left(\begin{array}{ccc}l_i&s_v&L_i\\m_{l_i}&m_{v_i}&-M_{L_i}\end{array}
\right)
\left(\begin{array}{ccc}L_i&s_c&I_i\\M_{L_i}&m_{c_i}&-M_i\end{array}\right)
\nonumber \\ &\times&
\left(\begin{array}{ccc}l_f&s_v&L_f\\m_{l_f}'&m_{v_i}&-M_{L_f}'\end{array}\right)
\left(\begin{array}{ccc}L_f&s_c&I_f\\M_{L_f}'&m_{c_i}&-M_f\end{array}\right)
\nonumber \\ &\times&
\left(\begin{array}{ccc}l_f&s_v&L_f\\m_{l_f}&m_{v_f}&-M_{L_f}\end{array}\right)
\left(\begin{array}{ccc}L_f&s_c&I_f\\M_{L_f}&m_{c_f}&-M_f\end{array}\right)
\nonumber \\ &\times&
\left(\begin{array}{ccc}l_i&l&l_f\\-m_{l_i}&m&m_{l_f}'\end{array}\right)
\langle(L_is_c)I_i\mid\mid M(El)\mid\mid (L_fs_c)I_f\rangle 
\nonumber \\ &\times&
Y^{\star}_{l_fm_{l_f}}(\hat{\vec k}).
\end{eqnarray}

The expression for the triple differential cross section for the above 
electromagnetic breakup process in the incident beam coordinate system (the 
spins are not observed) is given by 
\begin{eqnarray}
{d^3\sigma \over d\Omega _{cv}d\Omega _{cm}dE_{cv}} = {d\sigma ^{el}
\over d\Omega _{cm}}\rho(E_{cv})P(\hat{\bf k}),
\end{eqnarray}
where $d\sigma ^{el} \over d\Omega _{cm}$ is the Rutherford scattering 
cross section and $\rho(E_{cv})$ is the density of final states at relative 
energy $E_{cv}$. 
$P(\hat{\bf k})$, obtained from the matrix element in Eq.(6) above, 
is associated with the probability that the two fragments are scattered in
the final channel with relative motion wave vector $\vec k$. We expand
it in terms of the spherical harmonic $Y_{LM}(\hat {\bf k})$. It is given by
\begin{eqnarray}
P(\hat {\bf k})= \sum _{LM}A_{LM}Y_{LM}(\hat {\bf k}),
\end{eqnarray}
where
\begin{eqnarray}
A_{LM}&=&\left({4\pi Z_Ae\over \hbar v}\right)^2
\sum _{lml'm'\atop L_iL_i'L_fL_f'}
\sum _{l_il_i'\atop{l_fl_f'\atop I_fI_f'}}
(-1)^{s_c + s_v -l_i -l_i'- L_i -L_i'}
\nonumber \\ &\times&
(-1)^{- I_f + I_f' - I_i +m}{\cal R}^{-(l + l')}
{(2I_f + 1)(2I_f' + 1)\over (2l + 1)(2l' + 1)}
\nonumber \\ &\times&
\sqrt{(2L_i + 1)(2L_i' + 1)}(2L_f + 1)(2L_f' + 1)
\nonumber \\ &\times&
\sqrt{(2l_f + 1)(2l_f' + 1)(2L + 1)\over 4\pi}
\nonumber \\ &\times&
[\sum _{nn'}D^l_{mn}(-{\pi \over 2},{\pi \over 2},{\pi - \theta_{cm}\over 2})
D^{l'\star}_{m'n'}(-{\pi \over 2},{\pi \over 2},{\pi - \theta_{cm}\over 2})
\nonumber \\ &\times&
Y_{ln}({\pi\over 2}, 0)Y_{l'n'}({\pi\over 2}, 0)I_{ln}(\theta_{cm}, \xi)
I_{l'n'}(\theta_{cm}, \xi)]
\nonumber \\ &\times&
\left(\begin{array}{ccc}l_f&l_f'&L\\0&0&0\end{array}\right)
\left(\begin{array}{ccc}l&l'&L\\m&-m'&-M\end{array}\right)
\left\{\begin{array}{ccc}l&l'&L\\I_f'&I_f&I_i\end{array}\right\}
\nonumber \\ &\times&
\left\{\begin{array}{ccc}l'&L_i'&L_f'\\s_v&l_f'&l_i'\end{array}\right\}
\left\{\begin{array}{ccc}l&L_i&L_f\\s_v&l_f&l_i\end{array}\right\}
\left\{\begin{array}{ccc}L_f&L_f'&L\\l_f'&l_f&s_v\end{array}\right\}
\nonumber \\ &\times&
\left\{\begin{array}{ccc}l&I_i&I_f\\s_c&L_f&L_i\end{array}\right\}
\left\{\begin{array}{ccc}l'&I_i&I_f'\\s_c&L_f'&L_i'\end{array}\right\}
\left\{\begin{array}{ccc}I_f&I_f'&L\\L_f'&L_f&s_c\end{array}\right\}
\nonumber \\ &\times&
\langle(L_is_c)I_i\mid\mid M(El)\mid\mid (L_fs_c)I_f\rangle 
\nonumber \\ &\times&
\langle(L_i's_c)I_i\mid\mid M(El')\mid\mid (L_f's_c)I_f'\rangle^{\star}. 
\end{eqnarray}
In Eq.(9), $\cal R$ is half the distance of closest approach in a head-on 
collision
and $\xi$ is the adiabaticity parameter, given as the ratio of the collision
time and the excitation time. $D^l_{mn}$'s are the Wigner $D$-functions
and $\theta_{cm}$ is the scattering angle of the c.m. of the 
projectile. $Z_A$
is the charge number of the target and $v$ is the projectile-target relative
velocity in the entrance channel. $I_{ln}(\theta_{cm}, \xi)$ is the classical
orbital integral in the focal system of the hyperbolic orbit of the projectile
\cite{alw}.

Through our formalism, we can account for pure dipole ($l = l' = 1$) and pure
quadrupole ($l = l' = 2$) transitions as well as mixed transitions ($l = 1,
l' = 2$ or $l = 2, l' = 1$) or dipole-quadrupole interference.
Also, it is possible to calculate very exclusive observables up to the level 
of the triple differential cross section. Previous calculations
on the breakup of $^8$B were done by assuming that the angular distribution of 
fragments is isotropic in the projectile rest frame \cite{bertt,rsi,ian}. This
approximation gives
\begin{eqnarray}
{d^3\sigma \over d\Omega _{cv}d\Omega _{cm}dE_{cv}} =
{1\over 4\pi}{d^2\sigma \over d\Omega _{cm}dE_{cv}}.
\end{eqnarray}

It should be noted that expression for similar triple differential cross 
section as in Eq.(7) was given by Baur and Weber in Ref.~\cite{baw}, through
which it is possible to account for $E1$ and $E2$ contributions separately
as well as their interference. However, in contrast to our work, the coupling 
scheme of angular momenta and spins followed by them does not allow the use of
the more detailed coupling scheme adopted here. In fact, the coupling scheme in 
Ref.~\cite{baw} is more restricted in the sense that the orbital angular
momentum of the valence particle with respect to the core is coupled to the 
sum of the spins of $c$ and $v$ both in the initial and the final channel.
 
We note that integration over the solid angle associated with the
relative motion of the fragments gives
\begin{eqnarray}
{d^2\sigma \over d\Omega _{cm}dE_{cv}} = \sqrt{4\pi}{d\sigma ^{el}
\over d\Omega _{cm}}\rho(E_{cv})A_{00},
\end{eqnarray}
which is free from the $E1 - E2$ interference term, since $A_{00}$ (with $L, M$
= 0) does not involve these terms. This is evident from the 3-j symbol
$\left(\begin{array}{ccc}l&l'&L\\m&-m'&-M\end{array}\right)$ occurring in
Eq.(9), because for $L$ = 0, $l$ and $l'$ must be the same. Thus, any cross 
section obtained from the above triple differential 
cross section (Eq.(7)) by integration with respect to the solid angle $\Omega
_{cv}$ is free from $E1 - E2$ interference. This result is independent of the 
structure models of $^8$B. Therefore, the analyses presented in 
Refs.~\cite{mot,kik,rsi,ian} are indeed free from the $E1 - E2$ interference 
effects.

\section{\bf STRUCTURE MODEL}

The calculations of the reduced matrix elements involved in Eq.(9) require the 
detailed 
knowledge of ground state as well as continuum structure of $^8$B, which is not
yet known with certainty. In our calculations, we adopt the single particle 
potential model (SPPM) for $^8$B \cite{esb}. It is to be noted that 
the matrix elements of the multipole operators enter directly into these 
calculations. Therefore, they are quite sensitive to the structure model of 
$^8$B.

Within the SPPM, the valence proton (with spin $1\over 2$) in $^8$B 
(with spin-parity 2$^+$) is assumed to move relative to an inert $^7$Be 
core (with intrinsic spin-parity 3/2$^-$) in a Coulomb field and a Woods-Saxon 
plus spin-orbit potential, with an adjustable depth 
$V_0(l(Ls_c)I)$ for the initial and each final channel: 
\begin{eqnarray}
V(r)=V_0(l(Ls_c)I)(1-F_{s.o.}({\vec l}.{\vec s}){r_0\over r}
{d\over dr})f(r),
\end{eqnarray}
where
\begin{eqnarray}
f(r)=(1+\exp((r-R)/a))^{-1}.
\end{eqnarray}
Adjusting the depth allows one to reproduce the energy of the known states. We
use $a$=0.52 fm, $r_0$=1.25 fm and $R$=2.391 fm \cite{esb}. The spin-orbit 
strength is set to $F_{s.o.}$ = 0.351 fm \cite{esb}. The rms distance of the 
core-proton relative motion and the rms size of $^8$B come out to be 4.24 fm 
and 2.64 fm respectively.

The well depth for the ground state channel, $V_0(l_i(L_is_c)I_i)$ = $V_0((p
_{3/2}, 3/2^-)2^+)$, was adjusted to reproduce the one-proton separation energy
of 0.137 MeV. It came out to be $-$44.658 MeV). Similarly, the observed $I_f$ = 1$^+$ and
3$^+$ resonances in $^8$B are described as $p_{3/2}$ waves coupled to the 
ground state of the core, and the well depths for these channels, $-$42.14 and
$-$36.80 MeV, respectively, have been adjusted to reproduce the known resonance
energies (0.637 and 2.183 MeV respectively)\cite{ajz}. A $p_{3/2}$ wave and the 
spin of the core can also couple to
the total spin 0$^+$. But we ignore this channel since it appears to
be very weak in the low-lying excitation spectrum of $^8$B. For all other 
partial waves ($s_{1/2}, p_{1/2}, d_{3/2}$ etc.) we choose identical well depths
and set them equal to the value $-$42.14 MeV obtained for the 
$(p_{3/2}, 3/2^-)1^+$ channel, as suggested by Robertson \cite{rob}.

\section{\bf RESULTS and DISCUSSIONS}

To check the accuracy of our formulation, we present in Fig. 1 a comparison
of our calculations with the data \cite{kie} for the triple differential cross
sections for the reaction $^6$Li + Pb $\rightarrow$ $\alpha$ + $d$ + Pb
at 156 MeV beam energy as a function of the relative energy between the alpha
particle and the deuteron for $\theta _{\alpha} = \theta _d = $3$^{\circ}$. 
Only $E2$ excitation contributes in this case. Since
the multipole charge for the dipole case is zero for $^6$Li (Eq.(5)), the $E1$
contribution is zero. We
have assumed $^6$Li to be a cluster of $\alpha$ particle and deuteron, for which
the structure model 
\begin{figure}[here]
\begin{center}
\epsfxsize=8.0cm
\epsfbox{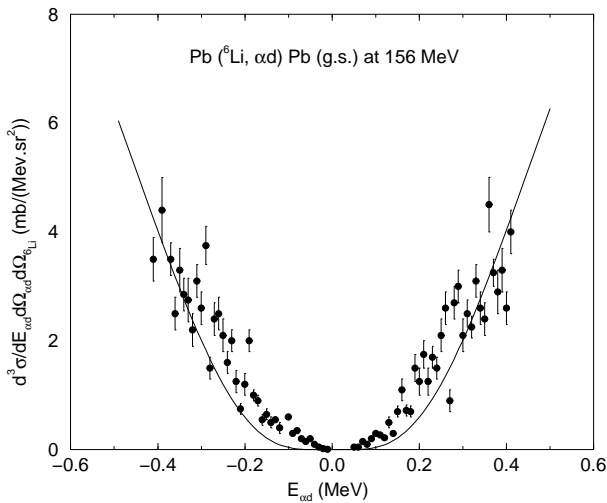}
\end{center}
\caption {
Triple differential cross section 
$d^3\sigma \over d\Omega _{\alpha d}dE_{\alpha d}d\Omega _{6_{Li}}$ for Coulomb 
breakup of $^6$Li on Pb at 156 MeV beam energy as a function of $E_{\alpha d}$ 
for $\theta _{6_{Li}}$ = 3$^{\circ}$ and $\theta _{\alpha d}$ = 0$^{\circ}$.
The data have been taken from \protect\cite{kie}. Positive $E_{\alpha d}$ 
implies velocity of $\alpha$ particle is larger than that of deuteron, while
negative $E_{\alpha d}$ corresponds to larger deuteron velocity than that of
$\alpha$ particle.
} 
\label{fig:figa}
\end{figure}
\noindent of Ref.~\cite{nis} has been used. We note that our 
calculations are in good agreement with the data. It may be noted that for the
results presented in Ref.~\cite{baw} for the same reaction, the structure part 
has been obtained from
a constant astrophysical $S$-factor of 1.7$\times$10$^{-5}$ MeV.mb, and not by
using proper wave functions for the ground state and excited states of $^6$Li
as has been done by us. In these data the contributions
from nuclear excitation effects are negligible as the freagments have been
detected at very forward angles \cite{shyam91}. 

In Fig. 2, we present the results of our calculations for the triple 
differential cross section (Eq.(7)) for Coulomb breakup of $^8$B on a Pb target
at 46.5 MeV/nucleon beam energy as a function of $\theta _{cm}$ for relative 
energy $E_{cv}$ = 0.2 MeV and $\theta _{cv}$ = 1$^{\circ}$ (top half) and as a 
function of $E_{cv}$ for $\theta _{cm}$ = 3$^{\circ}$ and $\theta _{cv}$ = 
1$^{\circ}$ (bottom half). We see that the $E1-E2$ interference (long dashed 
line) contribution is quite important in both cases. In fact, it is larger 
than the $E2$ contribution (dashed line) and modifies the coherent sum (solid 
line) of $E1$ (dotted line) and $E2$ cross sections significantly. 
It may be noted that this type of triple differential cross sections have not 
yet been measured for reactions involving unstable radioactive nuclei 
(e.g. $^8$B). 
\begin{figure}[here]
\begin{center}
\epsfxsize=8.0cm
\epsfbox{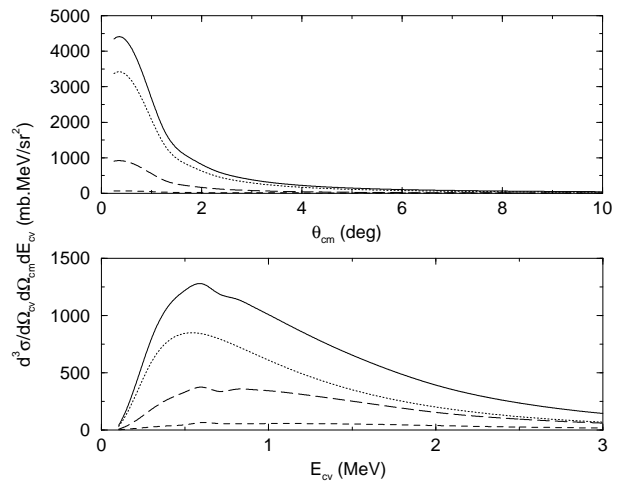}
\end{center}
\caption {
$E1$ (dotted line), $E2$ (dashed line) components together with the
$E1 - E2$ interference (long dashed line) contribution in the triple 
differential cross section 
${d^3\sigma \over d\Omega _{cv}d\Omega _{cm}dE_{cv}}$ for Coulomb 
breakup of $^8$B on Pb at 46.5 MeV/nucleon beam energy as a function of 
$\theta _{cm}$ for $E_{cv}$ = 0.2 MeV and $\theta _{cv}$ = 1$^{\circ}$ (top
half) and as a function of $E_{cv}$ for $\theta _{cm}$ = 3$^{\circ}$ and
$\theta _{cv}$ = 1$^{\circ}$ (bottom half). The solid line in each case shows 
the coherent sum of $E1$ and $E2$ cross sections.
} 
\label{fig:figb}
\end{figure}
However, the differential cross section $d\sigma \over d\theta _{
cm}$ has been measured as a function of $\theta _{cm}$ in the 
dissociation of $^8$B on Pb at beam energies $\sim$50 MeV/nucleon 
at RIKEN \cite{mot,kik}. At larger scattering angles  
the cross sections are more sensitive to the $E2$ component.
A detailed investigation of this reaction was carried out in Ref.~\cite{rsi} 
by assuming the fragment emission to be isotropic in the projectile rest
frame (Eq.(10)).

In Fig. 3, we present our calculations for the different components in the
Coulomb breakup cross section for this reaction at the beam energy of 51.9
MeV/nucleon, by using our formalism (i.e. without making the approximation of 
isotropic emission). For this observable 
the coherent sum (solid line) of $E1$ (dotted line) and $E2$ (dashed line) 
contributions is simply the sum of these two separate cross sections, as 
there is no contribution from the $E1 - E2$ interference component as discussed 
above (see Eq.(11)). We note that the results presented in Fig. 3 are the same 
as the pure Coulomb dissociation (with semiclassical theory) results presented 
in Ref.~\cite{plos}, which have been obtained with the isotropic emission 
assumption. These calculations, however, use a different structure model for
$^8$B, namely, the shell model embedded in the continuum. Therefore,
the present calculations give credence to the calculations reported in 
Ref.~\cite{plos}. As discussed earlier in Refs.~\cite{rsi,plos}, the 
disagreement between the data and the calculations beyond 4$^{\circ}$ in this
figure can be attributed to the point like projectile approximation of the
semiclassical theory, which is no longer valid for larger angles. Inclusion
of finite size effects in the calculations leads to better agreement between
data and the theoretical results \cite{rsi}. It should also be mentioned that
the nuclear breakup effects are important only at larger angles ($>4^\circ$)
in all the three energy bins \cite{rsi,bert}.
\begin{figure}[here]
\begin{center}
\epsfxsize=8.0cm
\epsfbox{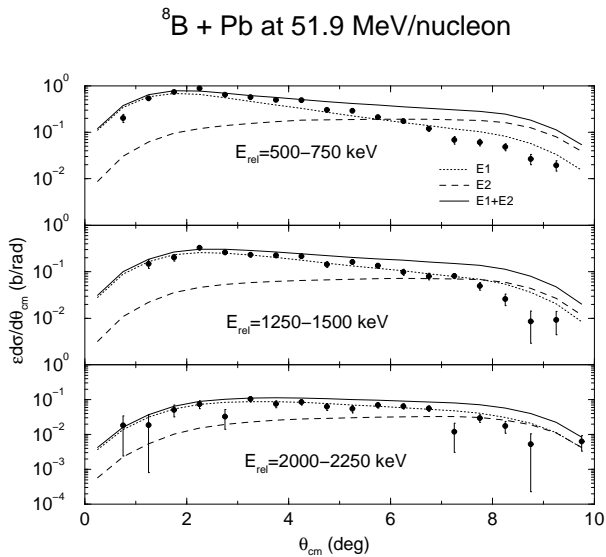}
\end{center}
\caption {
Calculated differential cross section
$\epsilon{d\sigma \over d\theta _{cm}}$ for Coulomb 
breakup of $^8$B on Pb at 51.9 MeV/nucleon for three given relative energy bins
as a function of $\theta _{cm}$. The cross sections have been folded with
experimental efficiency. The data, which are also folded with experimental
efficiency, have been taken from \protect\cite{kik}.
} 
\label{fig:figc}
\end{figure}
The triple differential cross section $d^3\sigma \over d\Omega _{cv}d\Omega 
_{cm}dE_{cv}$ can be related to that of the individual fragments $c$ and $v$
($d^3\sigma \over dE_cd\Omega _cd\Omega _v$) \cite{fuc}. Since interference
contributions are significant in $d^3\sigma \over d\Omega _{cv}d\Omega 
_{cm}dE_{cv}$, it is expected that they would also be important in  
$d^3\sigma \over dE_cd\Omega _cd\Omega _v$ in general. This is, indeed, the
case as can be seen in Fig. 4. In this figure, we have plotted 
$d^3\sigma \over dE_cd\Omega _cd\Omega _v$ (with $c \equiv {^7}Be$ and $v 
\equiv p$) as a function of $\theta _{7_{Be}}$ for Coulomb breakup of $^8$B 
on $^{58}$Ni at the beam energy of 25.8 MeV, for $E_{7_{Be}}$
= 22.575 MeV (the beam velocity energy), $\theta _p$ = 20$^{\circ}$ and
$\phi _p$ = 0$^{\circ}$. We see that the $E1 - E2$ interference term  
is quite significant and is even larger than that of the dipole component.
\begin{figure}[here]
\begin{center}
\epsfxsize=8.0cm
\epsfbox{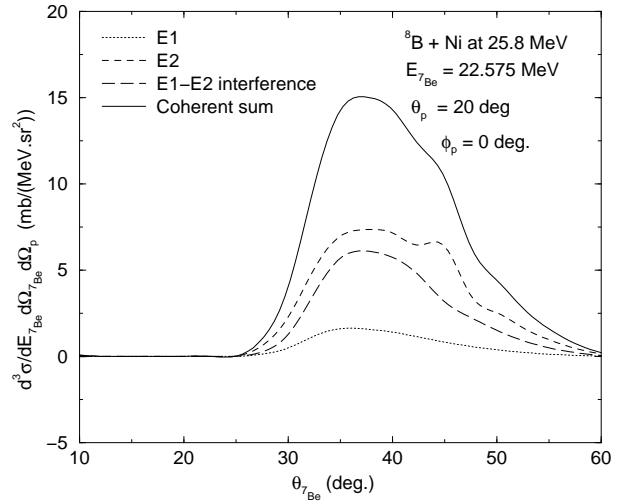}
\end{center}
\caption {
$E1$ (dotted line), $E2$ (dashed line) components together with the
$E1 - E2$ interference (long dashed line) contribution in the triple 
differential cross section 
${d^3\sigma \over dE_{7_{Be}}d\Omega _{7_{Be}}d\Omega _p}$ for Coulomb 
breakup of $^8$B on Ni at 25.8 MeV beam energy as a function of 
$\theta _{7_{Be}}$ for $E_{7_{Be}}$ = 22.575 MeV, $\theta _p$ = 20$^{\circ}$ 
and $\phi _p$ = 0$^{\circ}$. The solid line shows the coherent sum of $E1$ and 
$E2$ cross sections.
} 
\label{fig:figd}
\end{figure}
However, for less exclusive observables (like the angular distribution of
individual fragments) the interference term is not so strong. This can be
seen in Fig. 5, where we show the $^7$Be angular distribution resulting from 
the Coulomb dissociation of $^8$B on a Ni target at 25.8 MeV. We note that
the $E1 - E2$ interference (long dashed line) is small in magnitude 
compared to the dipole (dotted line) and quadrupole (dashed line) breakup cross
sections, excepting at larger angles (around 75$^{\circ}$) where it is 
comparable in magnitude to the $E1$ component. The interference 
component oscillates between positive and negative values which is also
reflected in the coherent sum of the $E1$ and $E2$ cross sections
(solid line).
\begin{figure}[here]
\begin{center}
\epsfxsize=6.7cm
\epsfbox{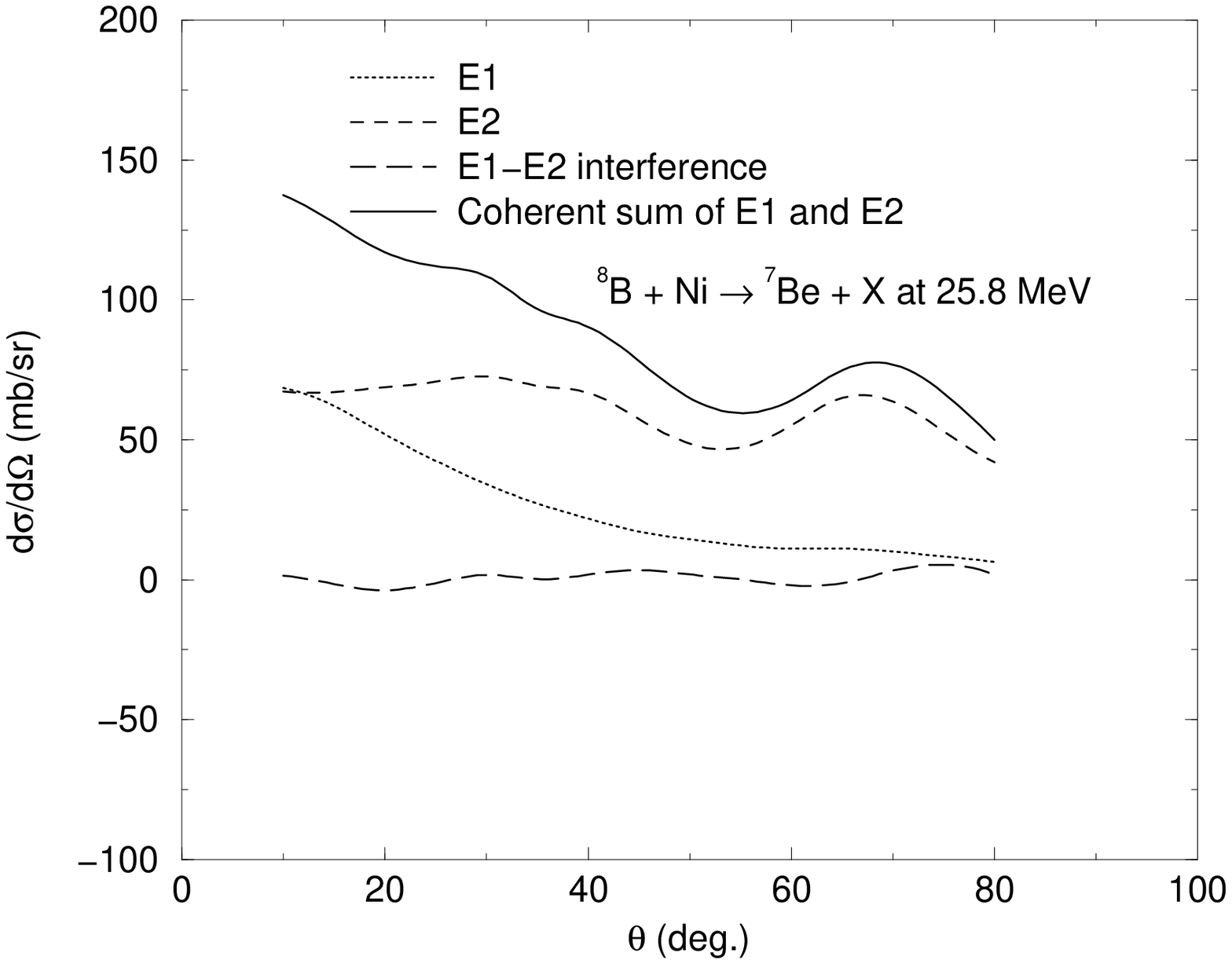}
\end{center}
\caption {
$E1$ (dotted line), $E2$ (dashed line) components together with the
$E1 - E2$ interference (long dashed line) contribution in the $^7$Be angular
distribution for Coulomb breakup of $^8$B on Ni at 25.8 MeV beam energy.
The solid line shows the coherent sum of $E1$ and $E2$ cross sections.
} 
\label{fig:fige}
\end{figure}
To understand why the interference is not large in the $^7$Be angular 
distribution shown in Fig. 5, we have plotted in Fig. 6 the $\theta _p$ (top 
half, with typical value of $\phi _p$ = 10$^{\circ}$) and $\phi _p$ (bottom 
half, with typical value of $\theta _p$ = 20$^{\circ}$) variations of the 
interference term in the triple differential cross
section shown in Fig. 4 for typical energy and angle of the $^7$Be fragment 
$E_{7_{Be}}$ = 22.575 MeV and $\theta _{7_{Be}}$ = 10$^{\circ}$ respectively.
The variation patterns are oscillatory and show positive and negative values
of almost equal magnitudes. For some angles the cross sections are very small. 
Therefore, it is no surprise that integrations with respect to the polar and 
azimuthal angles of the proton will lead to cancellation and hence to 
small magnitudes of the interference term in the overall $^7$Be angular 
distribution.
\begin{figure}[here]
\begin{center}
\epsfxsize=8.0cm
\epsfbox{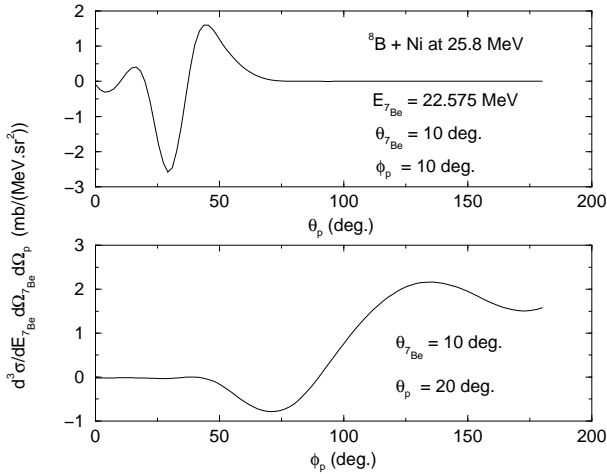}
\end{center}
\caption {
$\theta _p$ (top half, $\phi _p$ = 10$^{\circ}$) and $\phi _p$ (bottom 
half, $\theta _p$ = 20$^{\circ}$) variations in the
$E1 - E2$ interference contribution in the triple differential cross section 
${d^3\sigma \over dE_{7_{Be}}d\Omega _{7_{Be}}d\Omega _p}$ for Coulomb 
breakup of $^8$B on Ni at 25.8 MeV beam energy for $E_{7_{Be}}$ = 22.575 MeV 
and $\theta _{7_{Be}}$ = 10$^{\circ}$. 
} 
\label{fig:figf}
\end{figure}

However, there is still a remaining question of the effects of strong 
(nuclear) interaction between the colliding nuclei. They can
be almost eliminated by choosing the proper kinematical conditions.
For example, the $^8$B breakup data taken by the RIKEN collaboration 
\cite{mot,kik} have been shown \cite{rsi,bert,shyam}
to be  almost free from the nuclear effects for
$^7$Be-$p$ relative energies $\leq$ 0.75 MeV at very forward angles
($\leq$ 4$^\circ$). Similarly the data taken at GSI \cite{gsi} at higher
beam energy of $\sim$ 250 MeV/A are also free from these effects
(see ${\it e.g.}$ \cite{baur00}). Therefore, the results presented in 
Fig. 2 are unlikely to be affected by nuclear excitation. However,
these effects can be quite strong \cite{rsi} for the breakup studies  
\cite{von,gui} at lower beam energies ($\sim$ 25 MeV). In any case,
it is straightforward to calculate \cite{ribiki} the amplitudes for
pure nuclear and Coulomb-nuclear interference terms for the dipole
and quadrupole excitations as well as of their interference.

\section{\bf SUMMARY and CONCLUSIONS}

In summary, we have performed first order semiclassical calculations of the
electromagnetic excitation of $^8$B, where the
electric dipole and quadrupole excitation components as well as their 
interference effects are included. The theory permits consideration of 
coupling of angular momenta and spins of the projectile fragments in detail. 
It is possible to calculate exclusive observables to the level of the triple 
differential cross section within this formalism. 

We find that the magnitude of the $E1-E2$ interference term
could be appreciable 
in the triple differential cross sections. However, observables
which are not functions of the solid angle associated with
the relative motion of the breakup fragments in the final channel, are
free from this term. This result is independent of the structure
models of $^8$B. We have also shown that for double differential cross sections
involving angle of scattering of the projectile with respect to the target 
and energy of relative motion between the projectile fragments, the 
approximation of the isotropic emission in the rest frame of the projectile 
is quite good. Therefore, analysis of the RIKEN data presented earlier using
this approximation is quite accurate and this data, in the proper kinematical
regime ($\theta _{^8B^{\star}}<$4$^{\circ}$ and $E_{^7Be-p}\leq$500 keV) can
be used to extract rather reliable astrophysical $S$-factor $S_{17}$(0).

The interference terms are also significant in the triple differential cross
sections of the individual fragments. However, in the angular distribution
of the individual fragments these terms are not significant. Therefore, the
experimental data present in Ref.~\cite{gui} are almost free from the $E1
- E2$ interference terms. However, the effects of the three body kinematics 
(not considered in the analysis of these data) may still be important.

\end{document}